\documentclass[prl,aps,twocolumn,amsmath,footinbib,superscriptaddress]{revtex4-2}
\usepackage{amssymb}
\usepackage{graphicx}
\usepackage{hyperref}

\usepackage{color} 

\definecolor{darkgreen}{rgb}{0,0.7,0}  

\usepackage{subcaption}
\usepackage{ragged2e}
\usepackage{cancel}
\DeclareCaptionJustification{justified}{\justifying}
\usepackage{natbib}
\usepackage{caption}

\definecolor{orange}{RGB}{255,127,0}

\begin{document}

\title{Dynamical separation of charge and energy transport in one-dimensional Mott insulators}

\author{Frederik M{\o}ller}\email{frederik.moller@tuwien.ac.at}
\affiliation{
Vienna Center for Quantum Science and Technology (VCQ), Atominstitut, TU Wien, Vienna, Austria}
\author{Botond C. Nagy}\email{botond.nagy@edu.bme.hu}
\affiliation{Department of Theoretical Physics, Institute of Physics, Budapest University of Technology and Economics, H-1111 Budapest, M{\H u}egyetem rkp.~3.}
\affiliation{
BME-MTA Momentum Statistical Field Theory Research Group, Institute of Physics, Budapest University of Technology and Economics, H-1111 Budapest, M{\H u}egyetem rkp.~3.}
\author{M\'arton Kormos}\email{kormos.marton@ttk.bme.hu}
\affiliation{Department of Theoretical Physics, Institute of Physics, Budapest University of Technology and Economics, H-1111 Budapest, M{\H u}egyetem rkp.~3.}
\affiliation{MTA-BME Quantum Correlations Group (ELKH), Institute of Physics, Budapest University of Technology and Economics, H-1111 Budapest, M{\H u}egyetem rkp.~3.}
\affiliation{
BME-MTA Momentum Statistical Field Theory Research Group, Institute of Physics, Budapest University of Technology and Economics, H-1111 Budapest, M{\H u}egyetem rkp.~3.}
\author{G\'abor Tak\'acs}\email{takacs.gabor@ttk.bme.hu}
\affiliation{Department of Theoretical Physics, Institute of Physics, Budapest University of Technology and Economics, H-1111 Budapest, M{\H u}egyetem rkp.~3.}
\affiliation{MTA-BME Quantum Correlations Group (ELKH), Institute of Physics, Budapest University of Technology and Economics, H-1111 Budapest, M{\H u}egyetem rkp.~3.}
\affiliation{
BME-MTA Momentum Statistical Field Theory Research Group, Institute of Physics, Budapest University of Technology and Economics, H-1111 Budapest, M{\H u}egyetem rkp.~3.}

\date{\today}

\begin{abstract} 
One-dimensional Mott insulators can be described using the sine-Gordon model, an integrable quantum field theory that provides the low-energy effective description of several one-dimensional gapped condensed matter systems, including recent realizations with trapped ultra-cold atoms. Employing the theory of Generalized Hydrodynamics, we demonstrate that this model exhibits separation of the transport of topological charge vs.\ energy. Analysis of the quasiparticle dynamics reveals that the mechanism behind the separation is the reflective scattering between topologically charged kinks/antikinks.
The effect of these scattering events is most pronounced at strong coupling and low temperatures, where the distribution of quasiparticles is narrow compared to the reflective scattering amplitude. This effect results in a distinctively shaped ``arrowhead'' light cone for the topological charge.
\end{abstract} 

\maketitle 

\paragraph{Introduction.--- }One-dimensional (1D) quantum systems are well-known to exhibit anomalous transport behaviour compared with their higher-dimensional counterparts. In particular, transport in integrable quantum many-body systems \cite{2021RvMP...93b5003B} is strongly influenced by ergodicity breaking captured by the Mazur inequality \cite{1969Phy....43..533M,1971Phy....51..277S}, and it is primarily characterised by ballistic transport and finite Drude weights \cite{1995PhRvL..74..972C}. Another prominent anomaly is spin-charge separation, where the respective degrees of freedom in a one-dimensional quantum wire move with different velocities \cite{Giamarchi:743140}, as observed experimentally \cite{1999Natur.402..504S,2005Sci...308...88A,2006NatPh...2..397K,2009Sci...325..597J,2020Sci...367..186V,2022SciA....8M2781V}. This phenomenon is best understood in terms of bosonization leading to two Tomonaga-Luttinger liquids \cite{1950PThPh...5..544T,1963JMP.....4.1154L} with different speeds of sound. More recently, it was also understood directly in terms of the interacting Fermi gas~\cite{PhysRevLett.95.176401, PhysRevB.99.014305, PhysRevB.104.115423, 2022Science376.1305}. 

In this Letter, we demonstrate a similar, yet, at the same time, substantially different separation of energy and charge transport velocities by considering non-equilibrium dynamics in one-dimensional Mott insulators. Mott insulators are materials that are expected to be conducting based on conventional band theory; however, they fail to do so due to a gap induced by electron-electron interactions \cite{Mott:1990}. In the Tomonaga-Luttinger description of the charge sector of 1D systems, the gap is induced by Umklapp processes \cite{1991PhRvB..44.2905G}. These 1D Mott insulators include carbon nanotubes and organic conductors; their charge sector is described by the sine-Gordon field theory, which can be obtained via bosonization of the Hubbard model \cite{Controzzi2001}.

Besides electronic systems, the sine-Gordon field theory has numerous further applications ranging from spin chain materials~\cite{PhysRevB.79.184401, PhysRevLett.93.027201, PhysRevB.57.10592, PhysRevB.59.14376} through arrays of Josephson’s junctions~\cite{Lomdahl1985, PhysRevLett.55.2059} to trapped ultra-cold atoms \cite{PhysRevB.75.174511,2010PhRvL.105s0403C,2010Natur.466..597H,2017Natur.545..323S,Wybo2022,2023arXiv230316221W}, and can also be realized via quantum circuits \cite{2021NuPhB.96815445R} and coupled spin chains \cite{Wybo2022}. Recently, it was shown that the topological charge Drude weight in this model exhibits a fractal structure \cite{2023arXiv230515474N}, similar to that found for the spin Drude weight in the gapless XXZ spin chain~\cite{2013PhRvL.111e7203P, 2017PhRvL.119b0602I, 2019PhRvL.122o0605L, 2020PhRvB.101v4415A, 2022JPhA...55X4005I}. 

To study transport phenomena, we exploit the breakthrough of Ref.~\cite{2023arXiv230515474N}, which enabled applying Generalized Hydrodynamics (GHD)~\cite{2016PhRvX...6d1065C, 2016PhRvL.117t7201B} to the sine-Gordon model at generic values of the coupling. GHD gives access to the exact large-scale dynamics of integrable systems and has been immensely successful in numerous applications (see reviews~\cite{2021JSMTE2021k4004A, 2022JSMTE2022a4002D, Bouchoule_2022, Bastianello_2021, doyon2023generalized}), including the quantitative description of dynamics in several cold gas experiments~\cite{2019PhRvL.122i0601S, doi:10.1126/science.abf0147, PhysRevLett.126.090602, PhysRevX.12.041032}. Using GHD, we demonstrate that the dynamical separation of conserved quantities also occurs in the quantum sine-Gordon model in the form of topological charge and energy, as illustrated in Fig.~\ref{fig:illustration}. Similarly to the Fermi gas, the phenomenon follows from separate excitations, featuring different dispersion relations, being responsible for carrying the relevant quantities. However, a key difference from spin-charge separation is that energy-charge separation occurs in a gapped system. In addition, it also has a fractal structure analogous to the Drude weight when considered as a function of coupling.
Lastly, reflective scattering events can influence the ballistic transport of the topological charge in a peculiar fashion, which we demonstrate by considering a bump release protocol.

\begin{figure}
    \centering
    \includegraphics[width = 1\columnwidth]{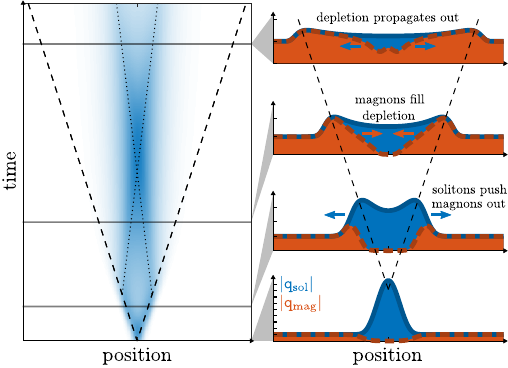}

    \caption{\label{fig:illustration}
    Illustration of the mechanism behind charge-energy separation and the three-staged ``arrowhead" light-cone propagation of the topological charge in the bump release:
    (\textit{i}) Outwards propagating solitons, whose front follows the dashed line, push all background magnons with them, following reflective kink/antikink scattering.
    (\textit{ii}) Magnons flow inwards to fill the depleted central region.
    (\textit{iii}) The remaining magnon depletion propagates outwards.
    The duration of each stage depends on the coupling strength and temperature.
    }
\end{figure}

\paragraph{Sine-Gordon hydrodynamics.---}
Sine-Gordon dynamics is driven by the Hamiltonian
\begin{equation}
    H=\int \mathrm{d} x\left[\frac{1}{2}\left(\partial_t \phi\right)^2+\frac{1}{2}\left(\partial_x \phi\right)^2-\lambda \cos (\beta \phi)\right] \text {, }
    \label{eq:SG_Hamiltonian}
\end{equation}
where $\phi(x)$ is a real scalar field, $\beta$ is the coupling strength, and the parameter $\lambda$ sets the mass scale.
The spectrum of the sine-Gordon model consists of topologically, and oppositely, charged \textit{kinks/antikinks} that are relativistic particles of mass $m_S$ interpolating between the degenerate vacua of the cosine potential.
In the repulsive regime $4 \pi < \beta^2 < 8 \pi$, kinks and antikinks comprise the entire spectrum, while in the attractive regime $0 < \beta^2 < 4 \pi$, kink-antikink pairs can form neutral bound states dubbed \textit{breathers}.
Introducing the renormalized coupling constant $\xi = \frac{\beta^2}{8 \pi - \beta^2}$, the breather masses are $m_{B_k} = 2 m_S \sin \left( \frac{k \pi \xi}{2} \right)$ where $k = 1, \ldots, n_B=\lfloor 1 / \xi\rfloor$.
For $\beta^2 > 8 \pi$, the cosine term of the Hamiltonian~\eqref{eq:SG_Hamiltonian} becomes irrelevant and the system reduces to the Luttinger liquid model.
We use units given by the kink mass $m_S$, $\hbar=1$ and the speed of light (the sound velocity in condensed matter context) $c=1$, as well as setting the Boltzmann constant $k_B=1$. 
As a result, energies and temperatures are measured in units of $m_S$, while distances and times are measured in units of $1/m_S$.


The root cause of the transport phenomenon lies in the dual nature of kink-antikink scattering, which can be both transmissive and reflective with respective amplitudes
\begin{align}
    S_T(\theta)&=\frac{\sinh \left( \theta /\xi \right)}{\sinh \left(( i \pi-\theta)/\xi\right)} S_0(\theta,\xi)\; , \\
    S_R(\theta)&=\frac{i \sin \left(\pi/\xi\right)}{\sinh \left((i \pi-\theta )/\xi\right)}  S_0(\theta,\xi)\; , \label{eq:reflective_amplitude}
\end{align}
where $\theta$ is the rapidity difference between the excitations and $S_0(\theta,\xi)$ is a phase factor. All other scattering processes are purely transmissive, with explicit expressions of their amplitudes given in the Supplemental Material~\cite{SM} (see also references~\cite{Bertini:2019lzy, inpreparation, Doyon_2020} therein).
For integer values of $1/\xi$, the kink-antikink reflection amplitude~\eqref{eq:reflective_amplitude} vanishes; at the aptly named reflection-less points of the coupling all topologically charged particles propagate at the same velocities, whereby any separation in transported quantities vanishes.

Thermodynamic states of the system can be described using the Bethe Ansatz~\cite{1969JMP....10.1115Y, takahashi_1999} and formulated in terms of quasiparticle excitations consisting of the breathers $B_k$, a single solitonic excitation $S$ accounting for the energy and momentum of the kinks, and also partly for the charge, and additional massless auxiliary excitations, dubbed \textit{magnons}, which account for the internal degeneracies related to the charge degrees of freedom of the kinks.
Whilst solitons carry a positive topological charge, magnons are negatively charged (see \cite{SM}).
The magnons can be classified by writing the coupling $\xi$ as a continued fraction
\begin{equation}
    \xi = \frac{1}{\displaystyle n_B+\frac{1}{\displaystyle\nu_1+\frac{1}{\displaystyle\nu_2 + \ldots}}}\ ,
\end{equation}
with $n_B$ breathers and $\nu_k$ magnon species at level $k$. The generic description of thermodynamic states was derived in \cite{2023arXiv230515474N}. It contains a set of equations of the overall form
\begin{equation}
    \rho_a^{\text{tot}}  = \eta_a s_a + \sum_b \eta_b \Phi_{ab}*\rho_b \, ,
    \label{eq:density}
\end{equation}
where the star denotes convolution, $\rho_a^{\text{tot}}(\theta)$ is the total density of states for excitations of type $a$ in rapidity space, $\rho_a(\theta)$ are the densities of occupied states, $\Phi_{ab}$ are kernels describing quasiparticle interactions, and $\eta_a$ are sign factors ensuring the positivity of the densities. The source terms $s_a=m_a \cosh\theta /2\pi$ contain the mass $m_a$ of the corresponding excitations, which is $m_S$ for solitons, $m_{B_k}$ for the $k$th breather and $m_a=0$ for magnons. The above equations only fix the relation between the total and occupied densities of states; in thermodynamic equilibrium, at temperature $T$ and chemical potential $\mu$ for the topological charge, all of them are fixed uniquely by the 
thermodynamic Bethe Ansatz (TBA) equations in terms of the pseudo-energy functions $\epsilon_a=\log\left({\rho_a^\text{tot}}/{\rho_a}-1\right)$
\begin{equation}
    \epsilon_a = w_a - \sum_{b}\eta_b \Phi_{ab}*\log\left(1+e^{-\epsilon_b}\right)\,,
    \label{TBA_struct}
\end{equation}
where the source terms are $w_a = m_a\cosh\theta/T - \mu q_a/T$ with $q_a$ giving the topological charge carried by the excitation of species $a$. More details, including the system's partially decoupled form and a graphical representation, can be found in \cite{2023arXiv230515474N,SM}.


\begin{figure}
    \centering
    \includegraphics[width = 1\columnwidth]{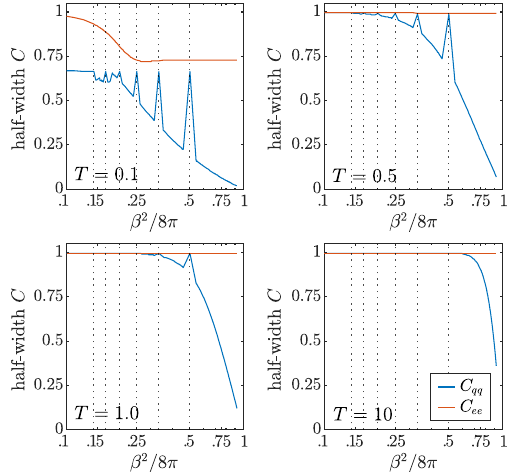}

    \caption{\label{fig:correlation_halfwidth}
    Half-width of the support of charge-charge (blue) and energy-energy (red) correlators in a bipartition protocol as function of the coupling strength $\beta^2 /8\pi$.
    The correlators are computed for dynamics at different temperatures and values of $\xi$ with at most two magnonic levels in the TBA system. The results are computed at discrete points, joined by a line in the plot to emphasise the discontinuous nature of the charge-charge case.
    The vertical dotted lines indicate the reflectionless points. Dimensionful quantities are given in units defined by setting $m_S=1$, $\hbar=1$ and $c=1$ as specified in the main text. 
    Note the logarithmic scale of the horizontal axis.
    }
\end{figure}

The large-scale dynamics of an inhomogeneous system can be expressed in terms of the evolution of the quasiparticle densities $\rho_a (z,t,\theta)$ via the theory of Generalized Hydrodynamics (GHD). In the absence of inhomogeneous couplings, the GHD equation reads~\cite{2016PhRvX...6d1065C, 2016PhRvL.117t7201B}
\begin{equation}
    \partial_t \rho_a (z,t,\theta) + \partial_z \left( v_a^{\mathrm{eff}}(z,t,\theta) \: \rho_a (z,t,\theta) \right) = 0  \; .
\end{equation}
We omit the $(z,t)$-dependence for a lighter notation in the following. The \textit{effective velocity} $v_a^{\mathrm{eff}}(\theta)$ represents the ballistic propagation velocity of a quasiparticle of type $a$ with rapidity $\theta$ and is given by 
\begin{equation}
    v_a^{\mathrm{eff}}(\theta)=\frac{\left(\partial_\theta e_a\right)^{\mathrm{dr}}(\theta)}{\left(\partial_\theta p_a\right)^{\mathrm{dr}}(\theta)} \; ,
\end{equation}
where $e_a (\theta) = m_a \cosh \theta$ is the bare energy of the quasiparticle type $a$ and $p_a (\theta) = m_a \sinh \theta$ is their bare momentum. The superscript `dr' indicates that the quantity has been \textit{dressed}, that is, it has been modified through interactions with other quasiparticles. As a result, the effective velocity carries an implicit dependence on the quasiparticle densities $\rho_a$ at the point $z$ and time $t$. The exact definition of the dressing operation and the TBA scattering kernels can be found in the Supplemental Material \cite{SM}. 
Physically, the effective velocity originates from the propagation of the quasiparticle excitations through the finite density medium \cite{PhysRevX.10.011054}; in the semi-classical picture, this modification can be understood as the accumulated effect of Wigner time delays associated with the phase shifts occurring under elastic collisions~\cite{PhysRevB.97.045407, PhysRevLett.120.045301}.

Finally, thermodynamic expectation values of local operators can be computed from the quasiparticle densities.
Thus, expectation values of densities of conserved quantities $\mathfrak{h}$ (such as topological charge and energy) are
\begin{equation}
    \langle \mathfrak{h}(z,t) \rangle \equiv \mathtt{h}(z,t) = \sum_a \int_{-\infty}^{\infty} \mathrm{d} \theta \: \rho_a (z,t,\theta) \: h_a (\theta) \; ,
\end{equation}
where $h_a(\theta)$ is the single-particle, \textit{bare} eigenvalue of the corresponding conserved quantity, such as $e_a(\theta)$ for the energy \cite{SM}.

\paragraph{Charge-energy separation.---}

In the limit of weak inhomogeneities, the separation of topological charge and energy follows from the different effective velocities of magnons and solitons.
To quantify the separation, we compute the charge-charge and energy-energy correlators at the hydrodynamic scale in thermal states, which indicate the maximal velocity of an energy or charge disturbance spreading on the thermal background, following~\cite{Doyon_2017, 10.21468/SciPostPhysCore.3.2.016}
\begin{equation}
\begin{aligned}
    &C_{\mathfrak{h}_1,\mathfrak{h}_2}(z,t) = \langle \mathfrak{h}_1(z,t) \mathfrak{h}_2(0,0)\rangle_c  \\
    &= t^{-1} \sum_a \sum_{\theta\in\theta_a^*(\zeta)} \frac{\rho_a(\theta) [1-\vartheta_a(\theta)]}{\left|\left(\partial_\theta v_a^{\text{eff}}\right)(\theta)\right|} h_{1,a}^{\text{dr}}(\theta)h_{2,a}^{\text{dr}}(\theta)\,,
\end{aligned}
\end{equation}
where $\zeta=z/t$, and $\theta_a^*(\zeta)$ are the set of rapidities for which the effective velocity takes the value $\zeta$, i.e., the solution of the equation $v_a^{\text{eff}}(\theta) = \zeta$.
The separation (and its absence) on the full range of the coupling $\beta^2/8\pi$ and for four different temperatures is shown in Fig.~\ref{fig:correlation_halfwidth}.
The figure depicts the half-width (in $\zeta$) of the correlators (see \cite{SM}).
It indicates that the separation strongly depends on the temperature in the attractive regime (where it is only visible at low temperatures), whilst it is more robust in the repulsive regime.
These dependencies follow from the relative rapidity width of the quasiparticle density to the reflective scattering amplitude; the former increasing with temperature, while the latter increases with coupling $\beta$.
Thus, in the repulsive regime, the amplitude $S_R(\theta)$ is generally wide compared to $\rho(\theta)$ up to high temperatures, whilst in the attractive regime, the width of $\rho(\theta)$ is comparable to $S_R(\theta)$ even at low temperatures.
In contrast, the kink-antikink scattering at reflectionless points is purely transmissive, whereby charge and energy propagate at the same velocity.
Notice the characteristic fractal structure in the dependence of the charge correlator half-width on the coupling, which is parallel to that found for the charge Drude weight in \cite{2023arXiv230515474N}.
Calculations of the half-width of topological charge- and energy-current profiles in a bipartition protocol with infinitesimal chemical potential and temperature differences of the two system halves reveal similar structures. For more details on the calculations for the bipartition protocol, see \cite{SM}.

\begin{figure}
    \centering
    \includegraphics[width = 1\columnwidth]{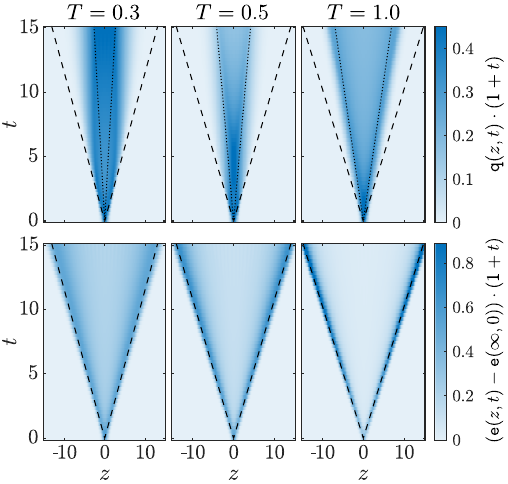}

    \caption{\label{fig:lightcones_repulsive}
    Evolution of topological charge density $\mathtt{q}$ and energy density $\mathtt{e}$ following a bump release in the repulsive $\xi = 3$ sine-Gordon model at three different temperatures $T$.
    Dashed and dotted lines indicate the position of the fastest travelling soliton and magnon for the background state, respectively.  
    The densities are scaled with the factor $(1+t)$ to emphasize features at later times. Dimensionful quantities are given in units defined by setting $m_S=1$, $\hbar=1$ and $c=1$ as specified in the main text. 
    }
\end{figure}

\begin{figure*}[]
    \centering
    \includegraphics[width = 1\textwidth]{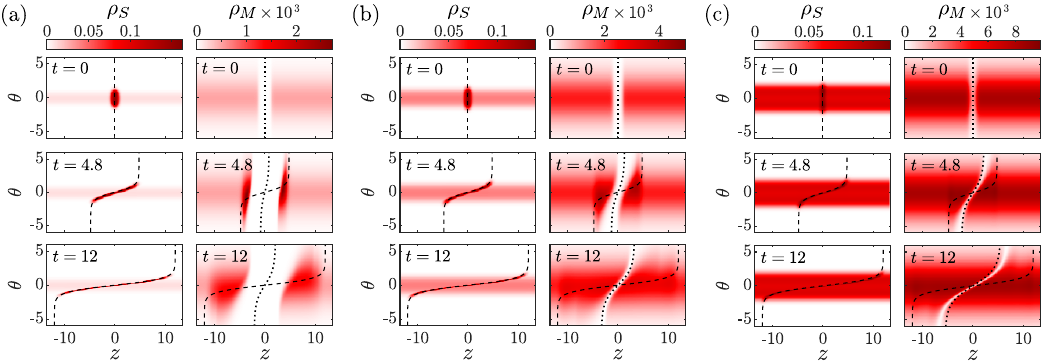}
    
    \caption{\label{fig:repulsive_quasiparticles} Soliton $\rho_S$ and (last) magnon $\rho_M$ distributions at different times $t$ following a bump release in the repulsive $\xi = 3$ sine-Gordon model for three temperatures: \textbf{(a)} $T = 0.3 $, \textbf{(b)} $T = 0.5 $, and \textbf{(c)} $T = 1.0 $. 
     The dashed and dotted lines indicate the positions $z = \overline v_{S}^{\mathrm{eff}}(\theta) t$ and $z = \overline v_M^{\mathrm{eff}}(\theta) t$, respectively. Dimensionful quantities are given in units defined by setting $m_S=1$, $\hbar=1$ and $c=1$ as specified in the main text. 
    }
\end{figure*}

\paragraph{``Arrowhead'' light-cone.---}
In the presence of strong inhomogeneities, reflective scattering events can lead to peculiar dynamics, which we demonstrate in a repulsive system with coupling $\xi = 3$, with one solitonic and $\nu_1 = 3$ magnonic excitation species.
The system is initialized in a local thermodynamic equilibrium at a given temperature $T$ and an inhomogeneous chemical potential profile $\mu(z)$, such that the initial topological charge density follows $q(z) = q_{\mathrm{max}} \exp\left( -\frac{z^2}{ 2 \sigma^2} \right)$, 
where $q_{\mathrm{max}} = 0.4$ and $\sigma = 0.5$.
This realizes a central region containing an excess of positively charged solitons and depletion of negatively charged magnons; in the charge-neutral background, their contribution is equal and opposite.
The dynamics is initiated by quenching the potential to zero at time $t = 0$.
Below we use $\overline v_{a}^{\mathrm{eff}}(\theta)$ to denote the effective velocity of quasiparticle species $a$ evaluated in the background state.
To simulate the GHD dynamics, we employ the backwards semi-Lagrangian method with a fourth-order scheme~\cite{10.21468/SciPostPhys.8.3.041, Moller2023}.

Fig.~\ref{fig:lightcones_repulsive} depicts the simulated charge and energy density evolution for temperatures $T = 0.3, \, 0.5, \, 1$.
For the energy density, a clear light cone is visible for all three temperatures, with higher temperatures featuring a sharper expansion profile.
The front of the light cone propagates with the velocity of the fastest solitons in the initial charge bump, indicated by the dashed line, which is obtained by first finding the endpoint of the rapidity interval containing 98\% of the soliton quasiparticles in the bump $\theta_{\mathrm{max}}$, then evaluating $\overline v_{S}^{\mathrm{eff}}(\theta_{\mathrm{max}})$.
The match between energy transport and soliton propagation is expected since only the solitonic excitations contribute to the energy.

In contrast, the evolution of the topological charge density exhibits a three-staged (``arrowhead'') light cone.
The mechanism behind this dynamics is illustrated in Fig.~\ref{fig:illustration}, while the underlying quasiparticle distribution is plotted at select times in Fig.~\ref{fig:repulsive_quasiparticles}~\footnote{Note, only the last (third) magnon species is plotted, as all species exhibit rather similar dynamics.}:
In the first stage, dynamics is dominated by the reflective scattering between kinks and antikinks; the energy-carrying solitons push all the background magnons with them, and the charge propagation matches the energy light cone.
The soliton propagation is hardly affected by interactions with the magnons.
This is evident from the soliton distribution of the initial bump dispersing according to their effective velocity in the background state $\overline v_{S}^{\mathrm{eff}}$, which is indicated by a dashed line in Fig.~\ref{fig:repulsive_quasiparticles}.
Meanwhile, for lower temperatures, the magnon propagation deviates strongly from their background velocity $\overline v_{M}^{\mathrm{eff}}(\theta)$ (plotted as a dotted line in Fig.~\ref{fig:repulsive_quasiparticles}), due to the magnons being pushed outwards by the expanding soliton bump.
The first stage lasts roughly until the charge contribution of the accumulated magnons cancels out that of the solitons; at this point, the outwards propagating charge front vanishes and magnons can propagate past the soliton front and start filling up the central depletion, thus shrinking the positively charged region.
In the final stage, as the inwards propagating magnons cross the center ($z=0$), a second outgoing light cone appears, effectively caused by the magnon depletion propagating outwards with velocity $\overline v_M^{\mathrm{eff}}$.

We find that the duration of the first and second stages exhibits a strong dependence on the temperature $T$.
For increasing temperature, the density of solitons and magnons in the background state grows, as seen in Fig.~\ref{fig:repulsive_quasiparticles}.
Thus, the point where the topological charge of the soliton front is cancelled by the accumulated magnon charge (marking the end of the first stage) is reached much sooner. 
In turn, this leads to the magnon depletion region being much narrower, thereby reducing the duration of the second stage.
Indeed, the charge propagation of the higher temperature realizations in Fig.~\ref{fig:lightcones_repulsive} follows almost solely stage three.
In the third stage, the initial, large perturbation has somewhat dispersed, whereby the system is only weakly inhomogeneous.
Thus, the charge-energy separation follows from the different effective velocities of magnons and solitons in thermal states; this difference decreases as $T$ increases, as the results shown in Fig.~\ref{fig:correlation_halfwidth} demonstrate. 

Additionally, we have simulated the bump release in the attractive regime; see \cite{SM} for figures depicting the results. Here, we find no clear ``arrowhead" structure in the charge propagation, as the different stages overlap. Similarly to the repulsive case, the dispersing soliton bump pushes magnons of the background state with it. However, as the rapidity width of the reflective scattering amplitude is much narrower in the attractive regime, the accumulated magnons can propagate past the solitons and fill the central magnon depletion immediately. Thus, the charge front of the propagating solitons is never (or at most only very slowly) cancelled by the magnon accumulation, whereby the first stage charge light cone (which follows the energy light cone) persists.

\paragraph{Summary.---} We uncovered a new effect of charge-energy separation in 1D Mott insulators, which manifests across a wide range of coupling strengths and temperatures using the framework of Generalized Hydrodynamics for the quantum sine-Gordon model. In the partitioning protocol, we have found that the separation exhibits a fractal structure similar to the Drude weight; at low temperatures, a clear separation is present at all coupling strengths except for the reflectionless points, while at higher temperatures and lower coupling strengths, the separation is suppressed. The bump release protocol sheds light on the underlying mechanism, which originates from the reflective part of the kink-antikink scattering. This mechanism implies that the effect is of a purely quantum origin and cannot be accounted for by the recent semiclassical approach to sine-Gordon GHD \cite{koch2023exact, bastianello2023sinegordon} since the classical scattering is purely transmissive. 
The role of reflective scattering is enhanced at low temperatures, especially in the repulsive regime, leading to a striking three-stage ``arrowhead'' light cone effect in the evolution of the topological charge.

The bump release, and similar protocols, can be experimentally realised by polarising the 1D Mott insulator via a locally applied voltage. Besides electronic systems, it can also be implemented in other realizations of sine-Gordon theory: for 1D magnets, the topological charge corresponds to spin, whereas the bump release can be realised using a locally applied magnetic field, while in cold atom systems, it can be achieved via a local shaping of the condensate as recently reported in \cite{2019OExpr..2733474T}.

\begin{acknowledgments}
\paragraph{Acknowledgements.} We thank Alvise Bastianello and Sebastian Erne for useful discussions. 
This work was supported by the National Research, Development and Innovation Office of Hungary (NKFIH) through the OTKA Grant ANN 142584. FM acknowledge support from the European Research Council: ERC-AdG: Emergence in Quantum Physics (EmQ) and partial support by the Austrian Science Fund (FWF) (Grant No. I6276). BN was partially supported by the Doctoral Excellence Fellowship Programme (DCEP) funded by the National Research Development and Innovation Fund of the Ministry of Culture and Innovation and the Budapest University of Technology and Economics, under a grant agreement with the National Research, Development and Innovation Office. GT was also partially supported by the NKFIH grant ``Quantum Information National Laboratory'' (Grant No. 2022-2.1.1-NL-2022-00004).
\end{acknowledgments}

\bibliographystyle{utphys}
\bibliography{references}

\clearpage
\setcounter{equation}{0}
\setcounter{figure}{0}
\setcounter{table}{0}
\setcounter{page}{1}
\makeatletter
\renewcommand{\theequation}{S.\arabic{equation}}
\renewcommand{\thefigure}{S.\arabic{figure}}
\appendix

\begin{widetext}
\section{SUPPLEMENTAL MATERIAL}
\subsection{Dynamical separation of charge and energy transport in one-dimensional Mott insulators}
\subsubsection{Frederik M{\o}ller, Botond C. Nagy, M\'arton Kormos, and G\'abor Tak\'acs}

\maketitle 

\section{The sine-Gordon TBA system}

In this section, we discuss the partially decoupled forms of the TBA equations (5,6), which makes the numerical calculation of the system possible by significantly reducing its computational complexity.

For brevity, here we only summarise the results for one magnonic level, i.e. when the coupling can be written as
\begin{equation}
    \xi = \frac{1}{n_B+\frac{1}{\nu_1}}\,.
\end{equation}
This case includes the repulsive regime $\xi=\nu_1\in \mathbb{Z}_{\geq 2}$ as well, by setting $n_B=0$, which was considered in \cite{Bertini:2019lzy}. For a full treatment valid for general values of the couplings, including both attractive and repulsive regimes, we refer to \cite{inpreparation}.

The TBA system consists of $n_B$ breathers, a soliton and $\nu_1$ magnons. The decoupled pseudo-energy system is
\begin{equation}
    \epsilon_a = \overline{w}_a + \sum_b K_{ab} * \left( \sigma_b^{(1)}\epsilon_b - \sigma_b^{(2)} \overline{w}_b + L_b \right)\,,
    \label{eq:TBA_decoupled}
\end{equation}
where $L_a=\log(1+e^{-\epsilon_a})$. The key advantage of this form is that the kernel $K_{ab}$ is a sparse matrix, as opposed to $\Phi_{ab}$ in Eq.(6). Note that the decoupling procedure modifies the source terms $w_a$ to $\overline{w}_a$. The modified source terms and the other constants appearing in Eqs.(\ref{eq:TBA_decoupled}, 6) are summarised in Table \ref{Table:TBA_constants}.
\begin{table}[b]
\centering
\begin{tabular}{|l|l|c|c|c|c|c|c|}
\hline
Excitations          & Labels                       & $\overline{w}$                                                 & $q$                  & $\overline{q}$ & $\eta$ & $\sigma^{(1)}$ & $\sigma^{(2)}$ \\ \hline\hline
Breathers            & $B_i$, $i=1,...,n_B$         & $m_{B_i}\cosh\theta/T$                                         & $0$                  & $0$            & $+1$   & $+1$           & $+1$           \\ \hline
Soliton              & $S$                          & $m_S \cosh\theta/T$                                            & $+1$                 & $0$            & $+1$   & $0$            & $0$            \\ \hline
Intermediate magnons & $M_{j}$, $j=1,...,\nu_1-2$   & 0                                                              & $-2 \cdot j$         & $0$            & $-1$   & $+1$           & $0$            \\ \hline
Next-to-last magnon  & $M_{\nu_1-1}$, $(j=\nu_1-1)$ & $\nu_1 \cdot \mu/T$                                            & $-2 \cdot (\nu_1-1)$ & $-\nu_1$       & $-1$   & $+1$           & $0$            \\ \hline
Last magnon          & $M_{\nu_1}$                  & $\nu_1 \cdot \mu/T$                                            & $-2$                 & $-\nu_1$       & $+1$   & $0$            & $0$            \\ \hline
\end{tabular}
\caption{The constants characterizing each species for a one magnon level TBA.}
\label{Table:TBA_constants}
\end{table}

The kernel $K_{ab}$ is most conveniently described in a graphical way, whereby the graphs encoding $K_{ab}$ consist of the building blocks summarised in Table~\ref{tab:Diagram_parts}.
\begin{table}[b]
\large 
    \centering
    \begin{tabular}{cccc}
        \includegraphics{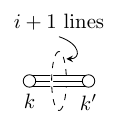} &
        \includegraphics{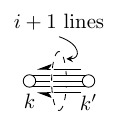} &
        \includegraphics{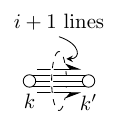} &
        \includegraphics{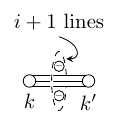} \\
        $K_{kk'}=K_{k'k}=\Phi_{p_i}$ & 
        $-K_{kk'}=K_{k'k}=\Phi_{p_i}$ & 
        $K_{kk'}=-K_{k'k}=\Phi_{p_i}$ & 
        $K_{kk'}=K_{k'k}=-\Phi_{p_i}$ \\
        &
        \includegraphics{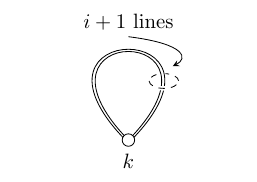} &
        \includegraphics{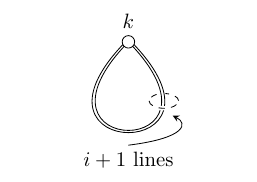} &
        \\
        &
        $K_{kk}=\Phi_{\text{self}}^{(i)}$ & 
        $K_{kk}=-\Phi_{\text{self}}^{(i)}$ &
    \end{tabular}
    \caption{Building blocks of diagrams encoding the sine-Gordon TBA systems at different couplings.}
    \label{tab:Diagram_parts}
\end{table}

The kernels can be written down analytically in Fourier space as
\begin{equation}
    \tilde{\Phi}_{p_i}(t) = \frac{1}{2\cosh\left(\frac{p_i}{\alpha}\frac{\pi}{2}\xi t\right)}\,, \quad \tilde{\Phi}_{\text{self}}^{(i)}(t) = \frac{\cosh\left(\frac{p_i-p_{i+1}}{\alpha}\frac{\pi}{2}\xi t\right)}{2 \cosh\left(\frac{p_i}{\alpha}\frac{\pi}{2}\xi t\right) \cosh\left(\frac{p_{i+1}}{\alpha}\frac{\pi}{2}\xi t\right)}\,,
    \label{eq:decoupledkernels}
\end{equation}
where
\begin{equation}
    \alpha=\nu_1\,, \quad p_0 = \nu_1\,, \quad p_1 = 1\,.
\end{equation}

With the above definitions, the $K_{ab}$ kernel in Eq.(\ref{eq:TBA_decoupled}) is encoded for one magnonic level as shown in Fig.~\ref{fig:one_level_graphs}.
\begin{figure}[t]
    \centering
    \begin{subfigure}{0.38\textwidth}
    \captionsetup{justification=centering}
        \includegraphics{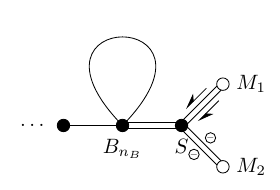}
        \caption{$\nu_1=2$.}
        \label{fig:tba_2magnon}
    \end{subfigure}
    \begin{subfigure}{0.58\textwidth}
    \captionsetup{justification=centering}
        \includegraphics{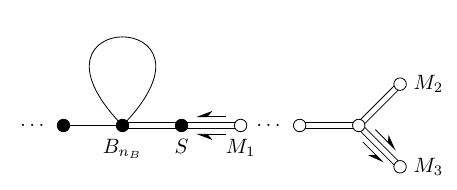}
        \caption{$\nu_1 \geq 3$.}
    \end{subfigure}
    \caption{Graphical representation of the kernel $K_{ab}$ in Eq.(\ref{eq:TBA_decoupled}) for couplings corresponding to one magnonic level.}
    \label{fig:one_level_graphs}
\end{figure}

For example, we spell out the kernel corresponding to Fig.~\ref{fig:tba_2magnon}.
\begin{equation}
    K_{ab} = 
    \begin{pmatrix}
        \ddots & \ddots & \vdots & \vdots & \vdots & \vdots\\
        \ddots  & 0 & \Phi_{p_0} & 0 & 0 & 0\\
        \dots  & \Phi_{p_0} & \Phi_{\text{self}}^{(0)} & \Phi_{p_1} & 0 & 0\\
        \dots  & 0 & \Phi_{p_1} & 0 & -\Phi_{p_1} & -\Phi_{p_1}\\
        \dots  & 0 & 0 & \Phi_{p_1} & 0 & 0\\
        \dots  & 0 & 0 & -\Phi_{p_1} & 0 & 0
    \end{pmatrix}\,,\quad\text{basis: }
    \begin{pmatrix}
        \vdots \\
        B_{n_B-1} \\
        B_{n_B} \\
        S \\
        m_1 \\
        m_2
    \end{pmatrix}\,.
\end{equation}

With the help of the pseudo-energies, one can calculate the ratio of occupied and all possible states, usually called the filling function
\begin{equation}
    \vartheta_a(\theta) = \frac{\rho_a(\theta)}{\rho_a^{\text{tot}}(\theta)} = \frac{1}{1+e^{\epsilon_a(\theta)}}\,.
\end{equation}
Elementary excitations modify the charges of finite density states by a different amount than their bare charges because the interactions with other particles dress up the bare charges. The dressed charges are given by
\begin{equation}
    \eta_a\,h_a^{\text{dr}} =  \overline{h}_a + \sum_b K_{ab}*\left[\left(\sigma_b^{(1)}-\vartheta_b\right)\eta_b\,h_b^{\text{dr}}-\sigma_b^{(2)} \overline{h}_b \right]\,.
    \label{eq:dressing_struct}
\end{equation}
In the dressing equation, the bare charges $h_i$ are again modified to $\overline{h}_a$ by the decoupling, at least for the topological charges $\overline{q}_a$, which are listed in Table \ref{Table:TBA_constants}. For the one-particle energy and momentum, the source terms aren't modified, i.e. $\overline{e}_a=e_a=m_a\cosh\theta$ and $\overline{p}_a=p_a=m_a\sinh\theta$. For the total densities of states, one can also show
\begin{equation}
    2\pi\rho_a^{\text{tot}} = \left(\partial_{\theta}p_a\right)^{\text{dr}}\,.
    \label{eq:total_dos}
\end{equation}
The bare velocities of each particle species are modified due to the scattering events with the sea of the other particles. For the resulting net propagation velocity, called the effective velocity, it can be shown
\begin{equation}
    v_a^{\text{eff}}(\theta) = \frac{\left(\partial_{\theta}e_a\right)^{\text{dr}}}{\left(\partial_{\theta}p_a\right)^{\text{dr}}}\,.
\end{equation}

\section{Kink-antikink scattering amplitudes}

Soliton scattering is described by the following two-particle amplitudes
\begin{equation}
\begin{aligned}
    S_{++}^{++}(\theta)&=S_{--}^{--}(\theta)=S_0(\theta)\,,\\
    S_{+-}^{+-}(\theta)&=S_{-+}^{-+}(\theta)=S_T(\theta)S_0(\theta)\,,\\
    S_{+-}^{-+}(\theta)&=S_{-+}^{+-}(\theta)=S_R(\theta)S_0(\theta)\, . 
\end{aligned}
\end{equation}
Here, +/- denotes kinks/antikinks, while $\theta$ is the difference in scattering rapidities.
Above, $S_T$ is the transmissive amplitude, $S_R$ is the reflective amplitude, and $S_0$ is the two-body scattering phase-shift, respectively defined as
\begin{equation}
\begin{aligned}
    S_T(\theta) &= \frac{\sinh\left(\frac{\theta}{\xi}\right)}{\sinh\left(\frac{i\pi-\theta}{\xi}\right)} \,, \\
    S_R(\theta) &= \frac{i\sin\left(\frac{\pi}{\xi}\right)}{\sinh\left(\frac{i\pi-\theta}{\xi}\right)}\,,\\
    S_0(\theta) &= -\exp\left(i\int\limits_{-\infty}^{\infty} \frac{\mathrm{d}t}{t}
                  \frac{\sinh\left(\frac{t\pi}{2}(\xi-1)\right)}{2\sinh\left(\frac{\pi\xi t}{2}\right)\cosh\left(\frac{\pi t}{2}\right)}{e}^{i\theta t}\right)\,.
\end{aligned}
\end{equation}
In figure~\ref{fig:scattering_amplitudes}, the reflective scattering probability $|S_R|^2$ is plotted as a function of rapidity for select coupling strengths $\xi$ in both the repulsive and attractive regimes. 
For increasing coupling strength, the width of $|S_R|^2 (\theta)$ increases; thus, in the repulsive regime, it is much wider than in the attractive one.
Note that for integer values of $1/\xi$, the kink-antikink reflection amplitude $S_R$ vanishes, corresponding to reflectionless (purely transmissive) scattering.

\begin{figure}
    \centering
    \includegraphics{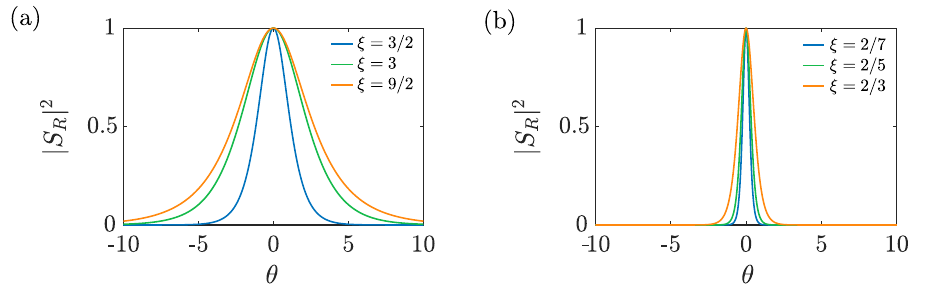}

    \caption{\label{fig:scattering_amplitudes}
    Reflective scattering probability $|S_R|^2$ in the \textbf{(a)} repulsive and \textbf{(b)} attractive regime as a function of scattering rapidity. The transmissive scattering probability is $|S_T (\theta)|^2  = 1 - |S_R (\theta)|^2 $. 
    }
\end{figure}

\section{Charge-energy separation in the bipartition protocol and from dynamical correlators}

The bipartition protocol is a very common protocol to study transport phenomena, where a system is cut in two halves, and the two sides are prepared in different states defined by the source terms
\begin{align}
    \overline{w}_i = 
    \begin{cases}
        \overline{w}_{i,L} &= \displaystyle\sum_h \beta^{(h)}_L \overline{h}_i\,,\quad z<0\,, \\
        \overline{w}_{i,R} &= \displaystyle\sum_h \beta^{(h)}_R \overline{h}_i\,,\quad z>0\,,
    \end{cases}
\end{align}
where $h$ are the conserved charges, i.e. the topological charge $q$, the momentum $p$ and the energy $e$, and possibly higher charges, and $\overline{h}$ are their values modified by the partial decoupling, while $\beta^{(h)}$ is the thermodynamically conjugate variable corresponding to $h$, e.g. $\beta^{(q)}=\mu/T$, $\beta^{(e)}=1/T$. After the system is let to evolve for an asymptotically long time, the state of the system is described by a filling function at each ray $\zeta=z/t$ \cite{2016PhRvL.117t7201B,2016PhRvX...6d1065C}
\begin{equation}
    \vartheta_i(\zeta, \theta) = \Theta\left(v_i^{\text{eff}}(\zeta,\theta)-\zeta\right)\vartheta_{i,L}(\theta) + \Theta\left(\zeta-v_i^{\text{eff}}(\zeta,\theta)\right)\vartheta_{i,R}(\theta)\,.
\end{equation}
Although this is an implicit equation for the fillings, as the effective velocities on the RHS depend on the filling, the usual recursive numerical scheme \cite{Doyon_2020, Bertini:2019lzy} quickly converges to the stable solution. The filling can then be used to calculate the total and the occupied densities of states through the dressing equations (\ref{eq:total_dos}) for each ray $\zeta$. Expectation values of charges and currents are then computed for each ray as
\begin{equation}
    \mathtt{h}(\zeta) = \sum_i \int \mathrm{d}\theta \rho_i(\zeta,\theta)  h_i(\theta) \,, \quad
    \mathtt{j}_{h}(\zeta) = \sum_i \int \mathrm{d}\theta \rho_i(\zeta,\theta) h_i(\theta) v^{\textrm{eff}}(\zeta,\theta)\,.
\end{equation}
Example current profiles obtained with the above prescription from the bipartition protocol for $\xi=3$, $T = 1$ and $\mu=0$ are shown in Fig.~\ref{fig:bipartition}. Note the half width depicted in the figures, which can be used to quantify the spreading velocity of energy and charge for the given temperature and coupling in Fig.~2.
\begin{figure}[t]
    \centering
    \includegraphics[height=6cm]{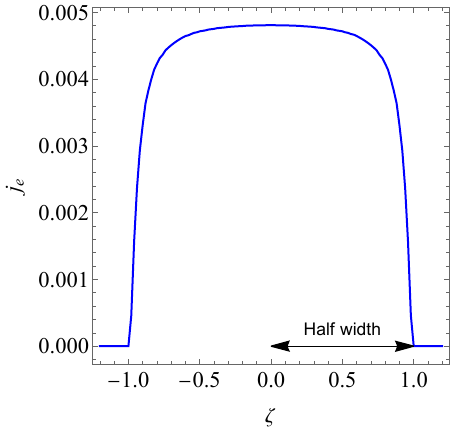}
    \hfil
    \includegraphics[height=6cm]{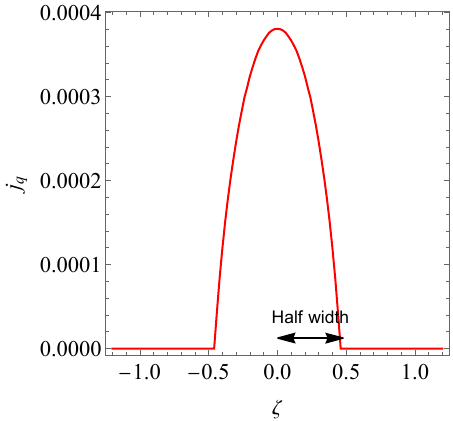}
    \caption{Energy (blue) and topological charge (red) current profiles in the bipartition protocol with infinitesimal chemical potential and temperature imbalance. Dimensionful quantities are given in units defined by setting $m_S=1$, $\hbar=1$ and $c=1$ as specified in the main text. }
    \label{fig:bipartition}
\end{figure}

Dynamical correlators describe how disturbances at one point in the system spread to distant points. It is possible to calculate the correlators of conserved charges in equilibrium states on the Euler scale in the TBA formalism as
\begin{equation}
    C_{h_1,h_2} (z,t) = \langle h_1(z,t) h_2(0,0)\rangle_c = t^{-1} \sum_a \sum_{\theta\in\theta_a^*(\zeta)} \frac{\rho_a(\theta) [1-\vartheta_a(\theta)]}{\left|\left(\partial_\theta v_a^{\text{eff}}\right)(\theta)\right|} h_{1,a}^{\text{dr}}(\theta)h_{2,a}^{\text{dr}}(\theta)\,,
\end{equation}
where $\zeta=z/t$, and $\theta_a^*(\zeta)$ are the set of rapidities for which the effective velocity takes the value $\zeta$, i.e. the solution of the equation $v_a^{\text{eff}}(\theta) = \zeta$. Examples of energy-energy and charge-charge correlators for $\xi=3$, $T = 1$ and $\mu=0$ are shown in Fig.~\ref{fig:correlators}. Note the half-width indicated in the figures, which are used to quantify the separation of the spreading of energy and charge in Fig.~2.
\begin{figure}[t]
    \centering
    \includegraphics[height=6cm]{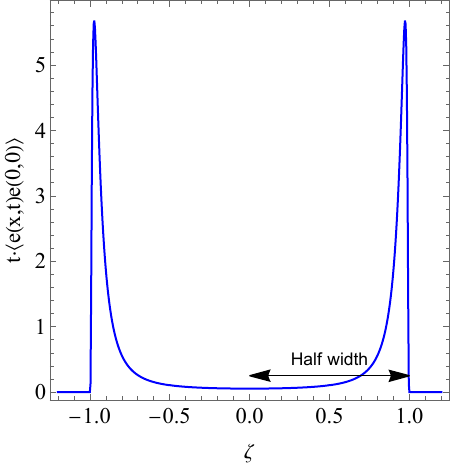}
    \hfil
    \includegraphics[height=6cm]{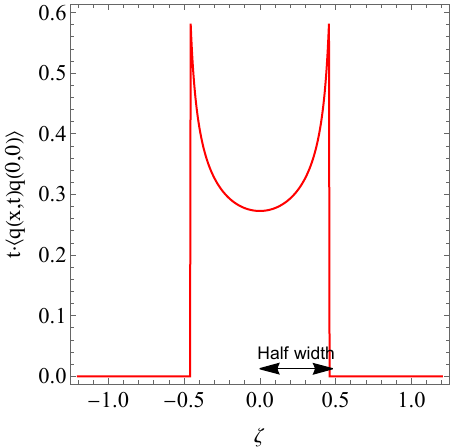}
    \caption{Dynamical energy-energy (blue) and charge-charge (red) correlators in equilibrium states. Dimensionful quantities are given in units defined by setting $m_S=1$, $\hbar=1$ and $c=1$ as specified in the main text. }
    \label{fig:correlators}
\end{figure}

\section{Additional figures from bump-release protocol}
In the following, we present several additional figures from the bump-release protocol discussed in the main text.
For the sake of convenience, we both summarize the contents of the figures and provide an analysis below:
\begin{itemize}
\item[Fig.~\ref{fig:profiles_repulsive}:] Profiles of topological charge density $\mathtt{q}$ and energy density $\mathtt{e}$ at select evolution times for $\xi = 3$ (repulsive interaction, same setup as discussed in main text). For $T = 0.3$, the charge dynamics follow stages 1 and 2 of the "arrowhead" light cone: At early times $t$, the charge-bump develops two peaks travelling outwards at the same velocity as the energy. Magnons accumulating at the soliton front eventually cancel out the charge contribution of the solitons, and a plateau in the charge density develops. During the second stage, the width of the plateau shrinks as the accumulated magnons fill the central depletion. At the very end of the second stage, the charge density profiles become peaked, as seen for the $T=0.5$ realisation around $t = 4.8$. Finally, a second outgoing light cone appears in stage three, following the magnon propagation velocity. 
\item[Fig.~\ref{fig:repulsive_velocities}:] Effective velocity evaluated at the right-moving soliton front at select evolution times for $\xi = 3$. Due to the excess of right-moving solitons, the magnon velocity is shifted to positive values following reflective kink/antikink scattering. As the soliton bump disperses and the local density of solitons and magnons becomes comparable, the shift in the magnon velocity decreases, and it tends towards its value in the background state (plotted in the top row).
\item[Fig.~\ref{fig:arrowhead}:] Bump-release at $\xi = 3$ for temperature $T=0.4$, clearly exhibiting all three stages of the "arrowhead" light-cone propagation. The figure shows \textbf{(a)} the (scaled) light cones of topological charge and energy density, \textbf{(b)} profiles (not scaled) of the charge and energy density at select evolution times, and \textbf{(c)} the quasiparticle distribution of solitons and (last) magnons at select times.
\item[Fig.~\ref{fig:lightcones_attractive}:] Light-cones of topological charge density and energy density following bump release for $\xi = 2/3$ (attractive regime). Unlike the repulsive case, no clear "arrowhead" structure is visible in the light cones; at all temperatures, the charge and energy propagations are very similar. As the main text explains, this follows from the much narrower reflective scattering amplitude in the attractive regime (see Fig.~\ref{fig:scattering_amplitudes}), which enables magnons to penetrate past the soliton front. Thus, a mixing of the first and second stages of the "arrowhead" dynamics occurs, whereby the dispersing soliton bump dominates both charge and energy transport.
\item[Fig.~\ref{fig:attrative_quasiparticles}:] quasiparticle distributions at select times for the $\xi = 2/3$ bump release. Comparing the soliton and magnon distributions, it is clear that the magnons are still experiencing significant "pushing" from the solitons. However, the effect is limited to rapidities around those of the solitons; indeed, for $z > 0$, the right-moving solitons (at positive rapidity) mainly push magnons also at positive rapidities. Meanwhile, magnons at negative rapidities continue to propagate inwards, filling up the initial depletion of magnons around $z=0$.
\item[Fig.~\ref{fig:attractive_velocities}:] Effective velocity evaluated at the right-moving soliton front at select evolution times for the $\xi = 2/3$ bump release. Similarly to the repulsive case, an excess of right-moving solitons causes a positive shift of the magnon velocity following reflective kink/antikink scattering. However, unlike the repulsive case, the effect is limited to mainly positive rapidities, while left-moving magnons at negative rapidities still exist (seen by the negative value of their effective velocity). As the soliton bump disperses and the local density of solitons and magnons becomes comparable, the shift in the magnon velocity decreases, and it tends towards its value in the background state (plotted in the top row). For temperatures around the soliton mass and greater, the velocity of solitons and magnons is practically identical.
\item[Fig.~\ref{fig:lightcones_reflectionless}:] Light-cones of topological charge density and energy density following bump release for $\xi = 1/3$ (reflectionless point). Following the absence of reflective scattering, no "arrowhead" structure is visible, and the charge and energy spread at the same rate throughout the system.
\item[Fig.~\ref{fig:reflectionless_quasiparticles}:] quasiparticle distributions at select times for the $\xi = 1/3$ bump release. In the absence of reflective scattering, the soliton distribution and the initial depletion of anti-solitons propagate in exactly the same manner.
\end{itemize}

\begin{figure}
    \centering
    \includegraphics{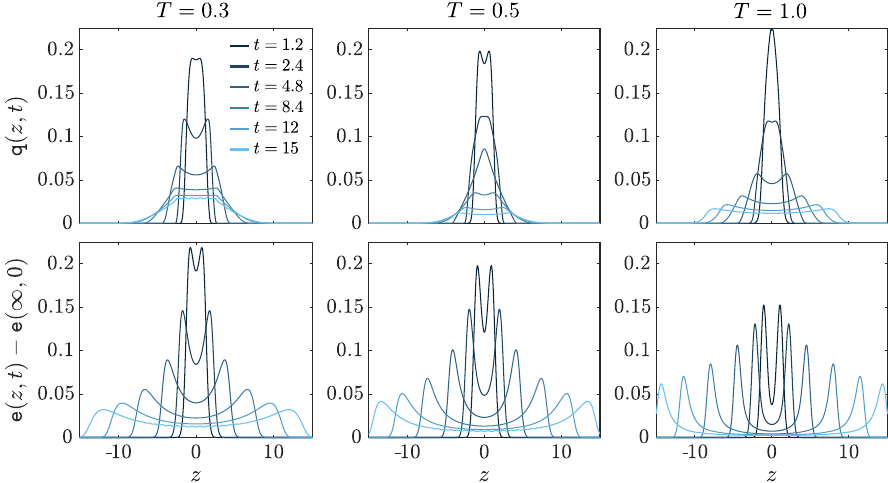}

    \caption{\label{fig:profiles_repulsive}
    Topological charge density $\mathtt{q}$ and energy density $\mathtt{e}$ following a bump release in the repulsive $\xi = 3$ sine-Gordon model at three different temperatures $T$ and at different evolution times $t$. Dimensionful quantities are given in units defined by setting $m_S=1$, $\hbar=1$ and $c=1$ as specified in the main text. 
    }
\end{figure}

\begin{figure}
    \centering
    \includegraphics{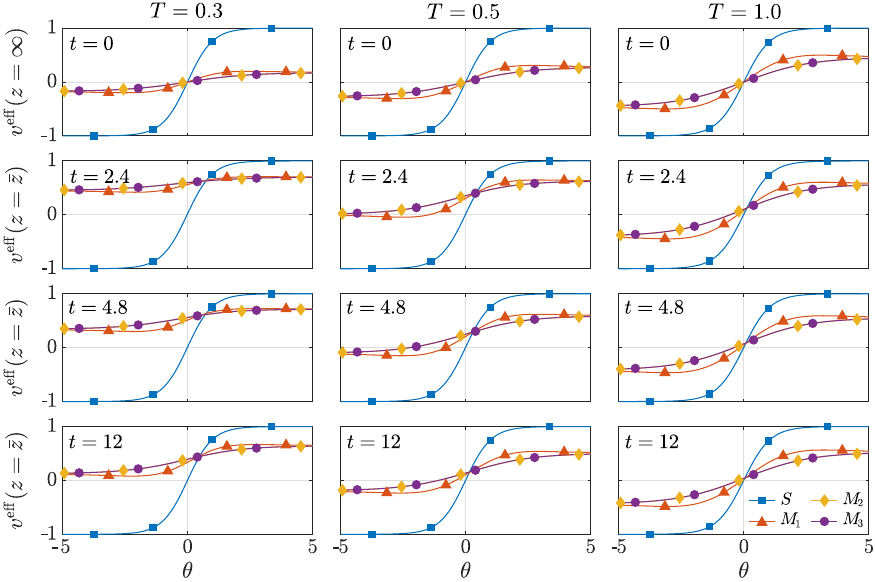}

    \caption{\label{fig:repulsive_velocities}
    Effective velocity $v_a^{\mathrm{eff}}$ for the four quasiparticle species found in the repulsive $\xi = 3$ sine-Gordon model, namely the soliton $S$ and three magnons $M_i$.
    Top row depicts the effective velocities evaluated for the background state in the bump release protocol for the three different temperatures $T$ (see main text).
    In the remaining rows, the effective velocity is evaluated at the position of soliton front $\bar z = \overline v_S^{\mathrm{eff}} t$ (see main text for definition) following different evolution times $t$ of the bump release.
    The presence of excess right-moving solitons, and thus an increase in reflective scattering events, manifests in the magnon effective velocity as a shift towards positive values. Dimensionful quantities are given in units defined by setting $m_S=1$, $\hbar=1$ and $c=1$ as specified in the main text. 
    }
\end{figure}

\begin{figure}
    \centering
    \includegraphics{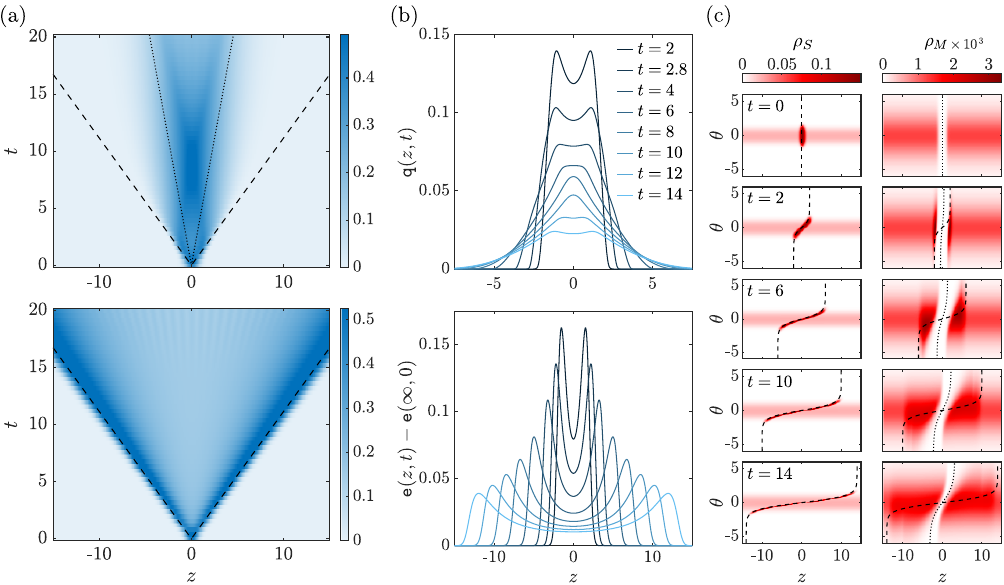}

    \caption{\label{fig:arrowhead}
    Dynamics following a bump release in the repulsive $\xi = 3$ sine-Gordon model at temperature $T = 0.4$.
    \textbf{(a)} Evolution of topological charge density $\mathtt{q}$ and energy density $\mathtt{e}$.
    Dashed and dotted lines indicate the position of the fastest travelling soliton and magnon for the background state, respectively.  
    The densities are scaled with the factor $(1+t)$ to emphasize features at later times.
    \textbf{(b)} Profiles of the topological charge density $\mathtt{q}$ and energy density $\mathtt{e}$ at select evolution times $t$.
    \textbf{(c)} Soliton and (last) magnon distribution at select evolution times. The dashed and dotted lines indicate the positions $z = \overline v_{S}^{\mathrm{eff}}(\theta) t$ and $z = \overline v_M^{\mathrm{eff}}(\theta) t$, respectively. Dimensionful quantities are given in units defined by setting $m_S=1$, $\hbar=1$ and $c=1$ as specified in the main text. 
    }
\end{figure}

\begin{figure}
    \centering
    \includegraphics{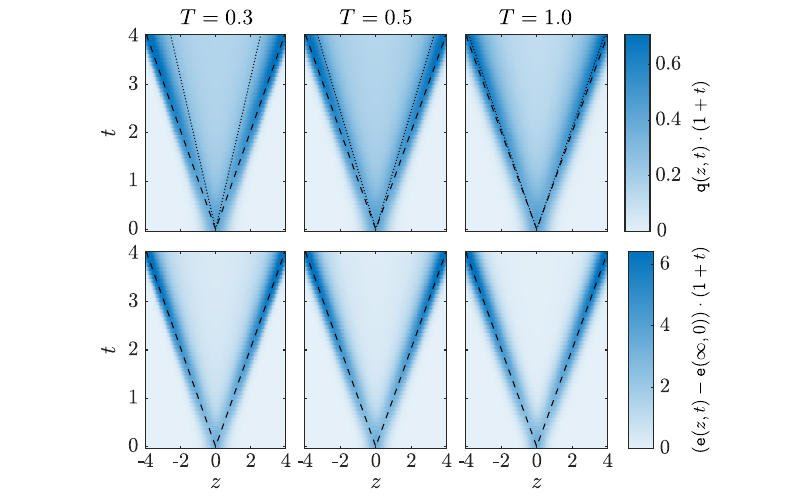}

    \caption{\label{fig:lightcones_attractive}
    The evolution of topological charge density $\mathtt{q}$ and energy density $\mathtt{e}$ following a bump release in the attractive $\xi = 2/3$ sine-Gordon model at three different temperatures $T$.
    Dashed and dotted lines indicate the position of the fastest travelling soliton and magnon for the background state, respectively.  
    The densities are scaled with the factor $(1+t)$ to emphasize features at later times. Dimensionful quantities are given in units defined by setting $m_S=1$, $\hbar=1$ and $c=1$ as specified in the main text. 
    }
\end{figure}

\begin{figure}
    \centering
    \includegraphics{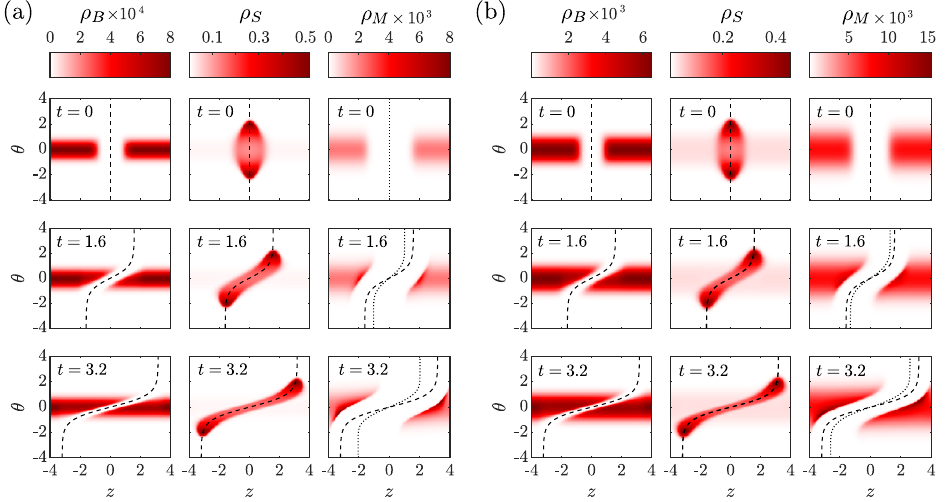}

    \caption{\label{fig:attrative_quasiparticles}
    Breather $\rho_B$, soliton $\rho_S$, and magnon $\rho_\nu$ distributions at different times $t$ following a bump release in the attractive $\xi = 2/3$ sine-Gordon model for two temperatures: \textbf{(a)} $T = 0.3 $ and \textbf{(b)} $T = 0.5$. 
    The dashed and dotted lines indicate the positions $z = \overline v_{S}^{\mathrm{eff}}(\theta) t$ and $z = \overline v_M^{\mathrm{eff}}(\theta) t$, respectively, where $\overline v_{a}^{\mathrm{eff}}(\theta)$ is the effective velocity of quasiparticle species $a$ evaluated with respect to the background state.
    Note that only the distribution of the first magnon species is plotted, as the dynamics of the other species is very similar. Dimensionful quantities are given in units defined by setting $m_S=1$, $\hbar=1$ and $c=1$ as specified in the main text. 
    }
\end{figure}

\begin{figure}
    \centering
    \includegraphics{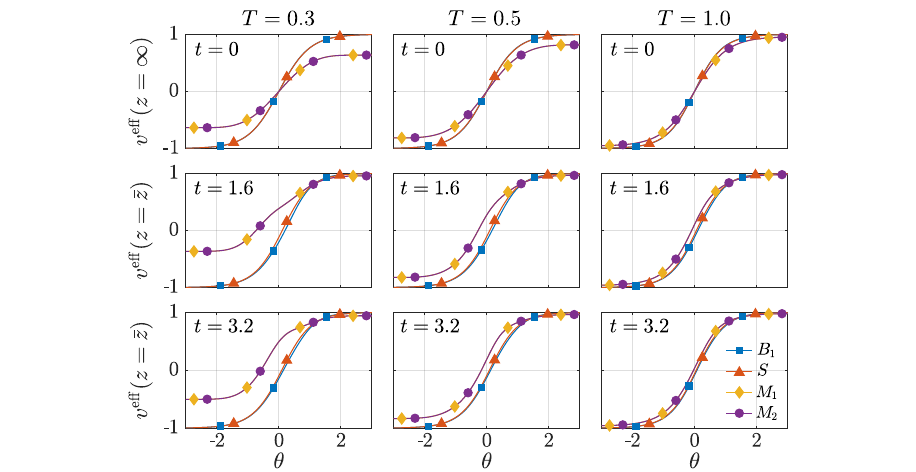}

    \caption{\label{fig:attractive_velocities}
    Effective velocity $v_a^{\mathrm{eff}}$ for the four quasiparticle species found in the attractive $\xi = 2/3$ sine-Gordon model, namely the breather $B$, soliton $S$ and two magnons $M_i$.
    The top row depicts the effective velocities evaluated for the background state in the bump release protocol for the three different temperatures $T$ (see main text).
    In the remaining rows, the effective velocity is evaluated at the position of soliton front $\bar z = \overline v_S^{\mathrm{eff}} t$ (see main text for definition) following different evolution times $t$ of the bump release.
    The presence of excess right-moving solitons, and thus an increase in reflective scattering events, manifests in the magnon effective velocity as a shift towards positive values. Dimensionful quantities are given in units defined by setting $m_S=1$, $\hbar=1$ and $c=1$ as specified in the main text. 
    }
\end{figure}

\begin{figure}
    \centering
    \includegraphics{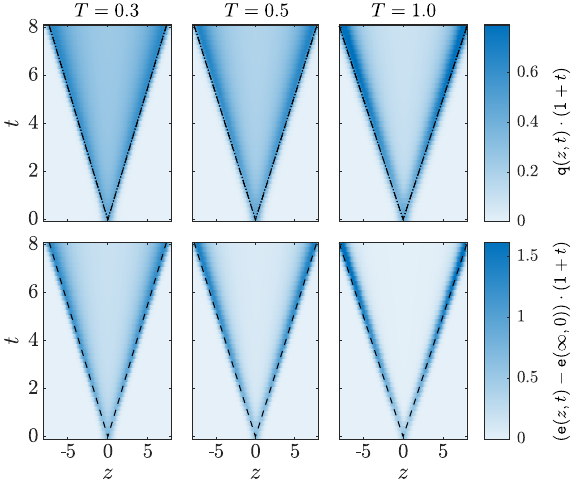}

    \caption{\label{fig:lightcones_reflectionless}
    Evolution of topological charge density $\mathtt{q}$ and energy density $\mathtt{e}$ following a bump release in the reflectionless $\xi = 1/3$ sine-Gordon model at three different temperatures $T$.
    Dashed and dotted lines indicate the position of the fastest travelling breather and soliton for the background state, respectively.  
    The densities are scaled with the factor $(1+t)$ to emphasize features at later times. Dimensionful quantities are given in units defined by setting $m_S=1$, $\hbar=1$ and $c=1$ as specified in the main text. 
    }
\end{figure}

\begin{figure}
    \centering
    \includegraphics{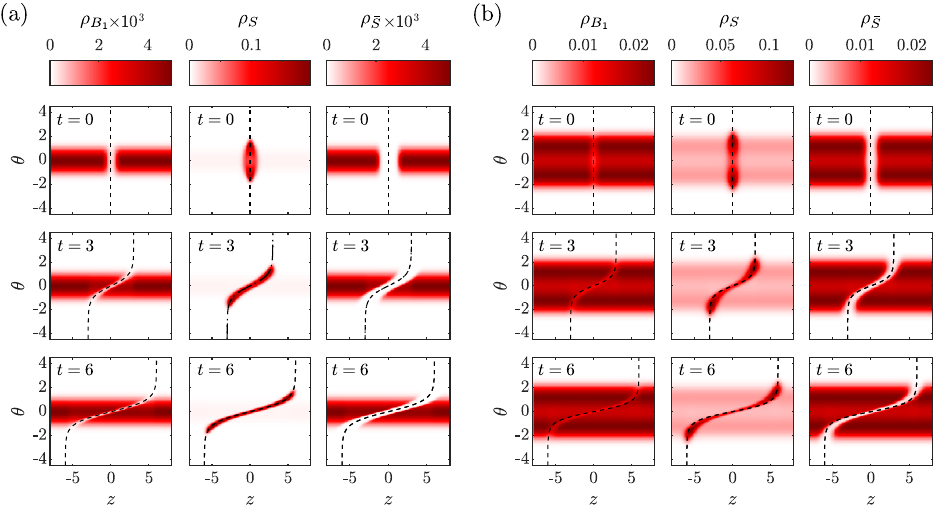}

    \caption{\label{fig:reflectionless_quasiparticles}
    Breather $\rho_B$, soliton $\rho_S$, and anti-soliton $\rho_{\bar S}$ distributions at different times $t$ following a bump release in the reflectionless $\xi = 1/3$ sine-Gordon model for two temperatures: \textbf{(a)} $T = 0.3 $ and \textbf{(b)} $T = 1.0 $. 
    The dashed lines indicate the positions $z = \overline v_{S}^{\mathrm{eff}}(\theta) t$, where $\overline v_{a}^{\mathrm{eff}}(\theta)$ is the effective velocity of quasiparticle species $a$ evaluated for the background state.
    Note that only the distribution of the first breather species is plotted, as the dynamics of the other species is very similar. Dimensionful quantities are given in units defined by setting $m_S=1$, $\hbar=1$ and $c=1$ as specified in the main text. 
    }
\end{figure}

\end{widetext}

\end{document}